\documentclass[aps,prl,singlecolumn,showpacs,superscriptaddress,groupedaddress,notitlepage]{revtex4-1}

\usepackage[english]{babel}
\usepackage{amsmath,amssymb}
\usepackage{graphicx}
\usepackage{dcolumn}
\usepackage{amsmath}
\usepackage{xcolor}

\usepackage{color}

\renewcommand{\phi}{\varphi}

\begin{document}
 


\title{Tangent map intermittency as an approximate analysis of intermittency in a high dimensional fully stochastic dynamical system: The Tangled Nature model}
 
\author{Alvaro Diaz-Ruelas}
\email{alvaropdr@fisica.unam.mx}
\affiliation{Instituto de F\'isica, Universidad Nacional Aut\'onoma de M\'exico, Ciudad Universitaria, Ciudad de M\'exico, 04510, Mexico} 
\author{Henrik Jeldtoft Jensen}
\email{h.jensen@imperial.ac.uk}
\affiliation{Centre for Complexity Science and Department of Mathematics, Imperial College London, South Kensington Campus, SW7 2AZ, UK}
\author{Duccio Piovani}
\email{duccio.piovani@gmail.com}
\affiliation{Centre for Complexity Science and Department of Mathematics, Imperial College London, South Kensington Campus, SW7 2AZ, UK \\Present address\\Centre for Advanced Spatial Analysis.
University College London, W1T 4TJ, London, UK}

\author{Alberto Robledo}
\email{robledo@fisica.unam.mx}
\affiliation{Instituto de F\'isica y Centro de Ciencias de la Complejidad, Universidad Nacional Aut\'onoma de M\'exico, Ciudad Universitaria, Ciudad de M\'exico, 04510, M\'exico}

\begin{abstract}
It is well known that low-dimensional nonlinear deterministic maps close to a tangent bifurcation exhibit intermittency and this circumstance has been exploited, \textit{e.g.} by Procaccia and Schuster [Phys. Rev. A 28, 1210 (1983)], to develop a general theory of 1/$f$ spectra. This suggests it is interesting to study the extent to which the behavior of a high-dimensional stochastic system can be described by such tangent maps. The Tangled Nature (TaNa) Model of evolutionary ecology is an ideal candidate for such a study, a significant model as it is capable of reproducing a broad range of the phenomenology of macroevolution and ecosystems. The TaNa model exhibits strong intermittency reminiscent of Punctuated Equilibrium and, like the fossil record of mass extinction, the intermittency in the model is found to be non-stationary, a feature typical of many complex systems. We derive a mean-field version for the evolution of the likelihood function controlling the reproduction of species and find a local map close to tangency. This mean-field map, by our own local approximation, is able to describe qualitatively only one episode of the intermittent dynamics of the full TaNa model. To complement this result we construct a complete nonlinear dynamical system model consisting of successive tangent bifurcations that generates time evolution patterns resembling those of the full TaNa model in macroscopic scales. In spite of the limitations of our approach, that entails a drastic collapse of degrees of freedom, the description of a high-dimensional model system in terms of a low-dimensional one appears to be illuminating.
\end{abstract}


\maketitle

\section{Introduction} 
High-dimensional complex systems, such as turbulence, relaxing glasses, biological evolution, the financial market or brain dynamics, exhibit intermittent dynamics \cite{Sibani2013,schuster1}. While intermittency in basic one-dimensional non-linear maps at the so-called tangent bifurcation \cite{schuster1} has received significant attention \textit{e.g.} because of their universal aspects \cite{Hu_Rudnick_1982} and has been suggested as a universal mechanism for 1/$f$ noise \cite{schuster2}. The relevance of such maps to high-dimensional stochastic systems depends on whether a robust macroscopic degree of freedom emerges, which is able to capture the dominant dynamics.  

A case in point is the Tangled Nature (TaNa) model \cite{Sibani2013} of evolutionary ecology, since it displays intermittent evolution at the macroscopic level while microscopically individuals reproduce, mutate and die at essentially constant rates \cite{tana:article1,tana:article2}. Numerical simulations of the model show that the total population $N(t)$ as a function of time $t$ (in the scale of generations) consists of quasi-stable, steady, periods that alternate with interludes of hectic transitions, during which $N(t)$ exhibits large amplitude fluctuations \cite{tana:article1,tana:article2}. The populations of species behave accordingly, during the quasi-stable periods they predominantly retain their identity, but at the transitions some species vanish, others arise, while the rest survive \cite{tana:article1,tana:article2}.

Here we study the incidence of intermittency, as displayed close to the tangent bifurcation in low-dimensional nonlinear maps, in the macroscopic behavior of the TaNa model. We make two intents. The first one is to approximate the evolution equations of the model, via determination of mean-field lowest-order local terms to obtain a map near tangency that reproduces the prototypical quasi-stable episode. The second is to model, phenomenologically, the sequences of consecutive quasi-stable and hectic periods via a nonlinear dynamical model that makes use of the families of tangent bifurcations that occur in one-dimensional quadratic maps.
      
We reach the conclusion that although the dynamics of the original TaNa model is fully stochastic and fluctuations are very important, the one-dimensional mean-field local map near tangency, we derive, and the consecutive tangent bifurcation model, we construct, do facilitate interesting insights that hint to a radical reduction of degrees of freedom under certain circumstances.

\section{The Tangled Nature model}
The Tangled Nature model is a model of evolutionary ecology, which studies the macro-dynamics emerging from the dynamics of individual organisms or agents, co-evolving together and subject to a web of mutual interactions. The model is an attempt to identify possible simple mechanisms behind the myriad of complicated interactions, feedback loops, contingencies, etc., as one moves from the short time reproductive dynamics at the level of individuals, to the long time systems level behaviour. The strategy is to keep the model sufficiently simple to enable analysis, and to pinpoint the details or assumptions in the model that are responsible for the specific behaviour at the systems level. One major concern of the model has been to understand how the smooth continuous pace of the reproductive dynamics at the level of individuals, can lead to intermittent or punctuated dynamics at the level of high taxonomic structures. The model was introduced in \cite{tana:article1,tana:article2} and since then, the model framework has been used by several authors see \textit{e.g.} \cite{FakeTanaBasic,FakeTaNaFluctuations,Becker_Sibani_2014,Nicholson_Sibani_2015,Vazquez_2015}. A summary of some of the models features and predictions can be found in \cite{Sibani2013}.    

\subsection*{Description of the model}
The dynamical entities of the TaNa model consist of  agents represented by a sequence of binary variables with fixed length $L$ \cite{Higgs/Derrida:article}. We denote by $n({\bf S}^a,t)$ the number of agents of type ${\bf S}^a=(S^a_1,S^a_2,...,S^a_L)$ (here $S^a_i\in\{-1,1\}$)at time $t$ and the total population is $N(t)=\sum_{a=1}^{2^L} n({\bf S}^{a},t)$. 
A time step is defined as a succession of one annihilation and of one reproduction attempt. Annihilation consists of choosing an agent at random with uniform probability and remove the agent with probability $p_{kill}$, taken to be constant in time and independent on the type. Reproduction: choose with uniform probability an agent, $\bf{S}^a$,  at random and duplicate the agent (and remove the mother) with probability
\begin{equation}
p_{kill}({\bf S}^a,t)=\frac{\exp{(H({\bf S}^a,t))}}{1+\exp{(H({\bf S}^a,t))}},
\label{eq:2.3}  
\end{equation}
which depends on the occupancy distribution of all the types at time $t$ through the weight function
 \begin{equation}
H({\bf S}^a,t)= \frac{k}{N(t)}\sum_b J({\bf S}^a,{\bf S}^b)n_b(t) -\mu N(t).
\label{eq:1}
\end{equation}
In Eq. (\ref{eq:1}), the first term couples the agent ${\bf S}^a$ to one of type ${\bf S}^b$ by introducing the interaction strength $\mathbf{J}({\bf S}^a,{\bf S}^b)$, whose values are randomly distributed in the interval $\left[-1,+1\right]$. For simplification and to emphasize interactions we here assume: $\mathbf{J}(\mathbf{S}^a,\mathbf{S}^a)=0$. The parameter $k$ scales  the interactions strength and $\mu$ can be thought of as the carrying capacity of the environment. An increase (decrease) in $\mu$ corresponds to harsher (more favourable) external conditions.

\begin{figure}[ht!]
\centering
\includegraphics[width=8cm,height=6.67cm]{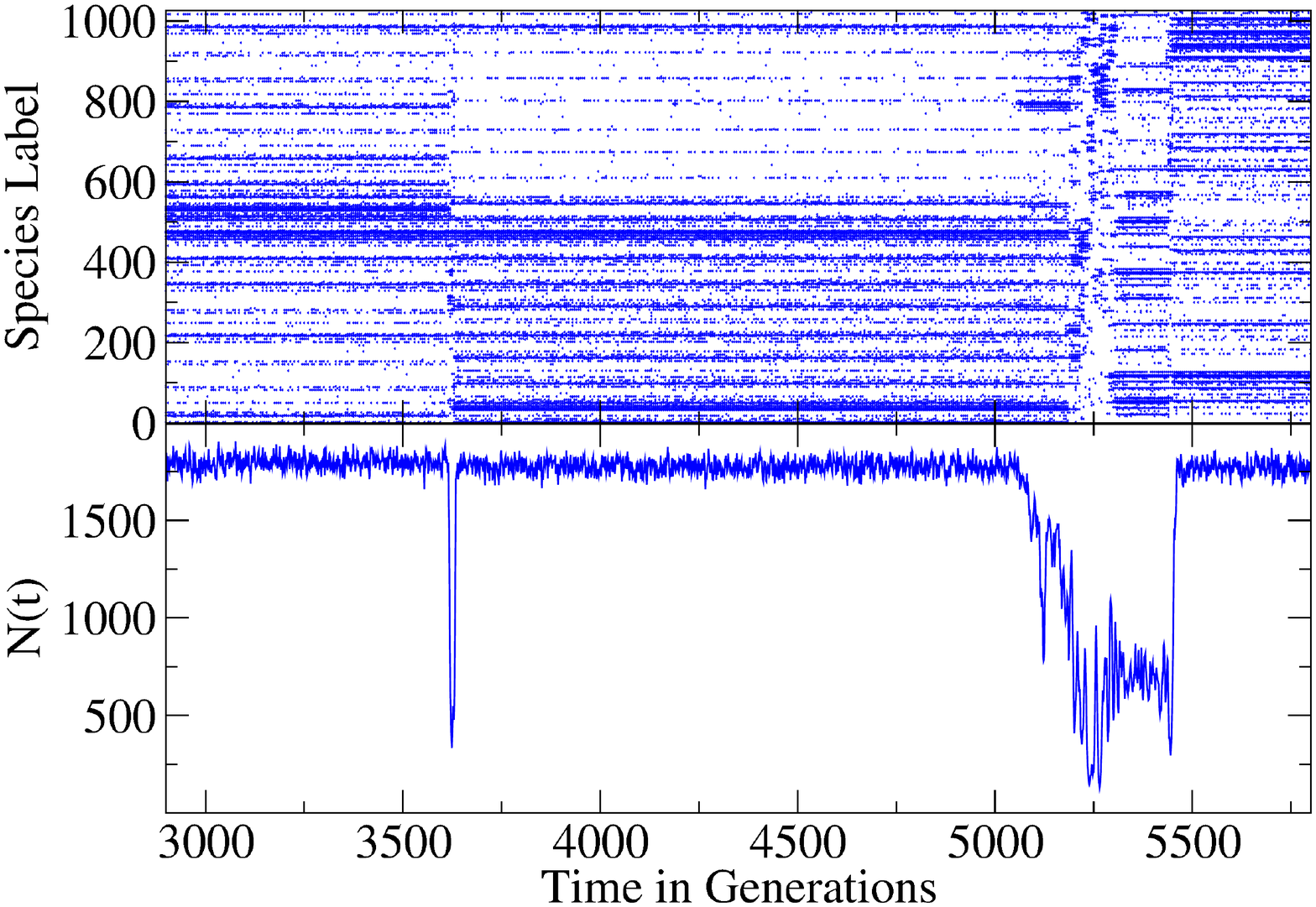}
\includegraphics[width=8cm,height=6.67cm]{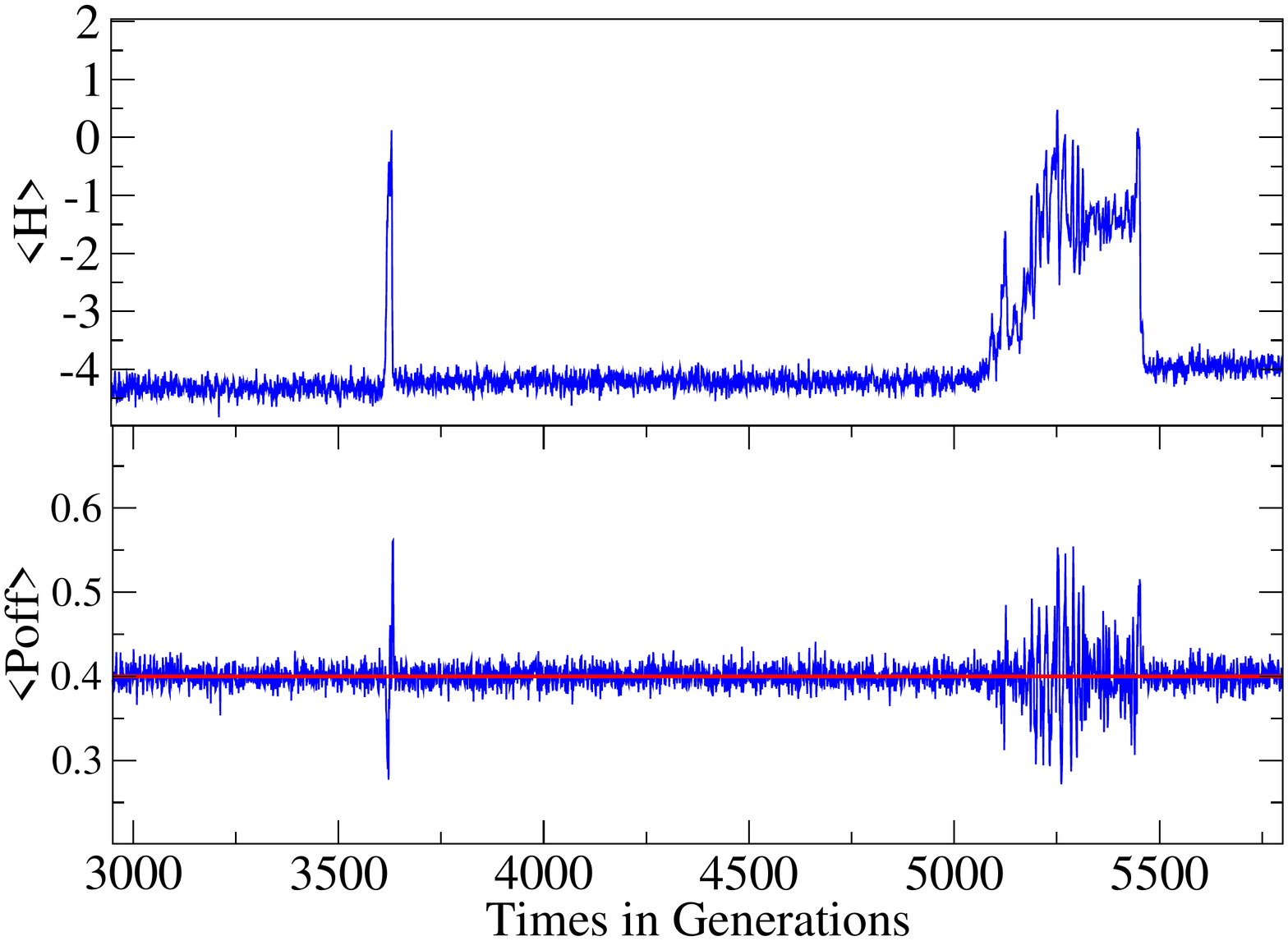}
\caption{Left Panel: Total population as a function of time (in generations) for a single realization of the TaNa model. The punctuated dynamics is clearly visible: quasi-stable periods alternate with periods of hectic transitions, during which $N(t)$ exhibits large amplitude fluctuations. Right panel: The average of the weight function $H$ and the reproduction probability. The parameters are $L = 10$, $p_{kill}  = 0.4$, $p_{mut} = 0.02$, $\mu  = 0.007$, $k = 40$ the red line indicates $p_{kill}$.}
\label{fig:1}
\end{figure}

Mutations occur in the following way: For each of the two copies ${\bf S}^{a_1}$ and ${\bf S}^{a_2}$, a single mutation changes the sign of one of the genes: $ S^{a_1}_i\rightarrow -S^{a_1}_{i}$, $ S^{a_2}_i\rightarrow -S^{a_2}_{i}$ with probability $p_{mut}$. We define a generation to consist of $N(t)/p_{kill}$ time steps, \textit{i.e.} the average time needed to kill all the individuals at time $t$. These microscopic rules generate intermittent macro dynamics\cite{tana:article2} as shown in Fig. \ref{fig:1}. The long quiescent epochs are called quasi Evolutionary Stable Strategies (qESS), since they do remind one of John Maynard Smith's notion of Evolutionary Stable Strategies  introduced in his game theoretic description of evolution \cite{Maynard:book}. 

The weight function $H$ will fluctuate about the value given by the stable dynamical fixed point condition $p_{off}(H)=p_{kill}$. This suggests that the mean field value of $H$ may indeed evolve in an intermittent way that may be captured by a tangent map. We will therefore derive the mean field map for $\langle H\rangle$. 
  
\subsection*{Derivation of Mean Field map for $H$}  
To establish a map for the mean field approximation to the weight function $H$, we need to analyse each of the microscopic stochastic processes that can lead to a change in $H$. These are reproduction, with or without mutation and death.  And we will make use of the fact that if a quantity, say $X$, undergoes the change to $X\mapsto X+\Delta$ with probability $p$ and remains unchanged $X\mapsto X$ with probability $1-p$, then in the mean field approximation we have $\langle X \rangle\mapsto \langle X \rangle +p\Delta$. We use a short hand notation in which we label individuals and types as $i,j,k...$ and accordingly the interaction between two types $i$ and $j$ as $J_{ij}$. \\

{\it Reproduction with no mutation}. We need to estimate the average change to the likelihood function, $H_i$ for type $i$ given that an individual of type $j_0$ reproduces without mutating. The change in $H_i$ is given by 
\begin{eqnarray}
H_i &\mapsto & \frac{k}{N+1} \left[\sum_{j\neq j_0} J_{i,j}n_j + J_{i,j_0}(n_{j_0} + 1)\right] - \mu(N+1) \nonumber \\
&=& \frac{k}{N+1} \left(\sum_{j}J_{i,j}n_{j}-\mu N\right) + \left(\frac{k}{N+1}J_{i,j_0}-\mu \right)\nonumber \\
&=& H_i + \Delta^i_{R,0m}(j_0)
\end{eqnarray}
We replaced $N+1$ by $N$ in the first term and introduced the change 
\begin{equation}
	\Delta^i_{R,0m}(j_0)= \frac{k}{N+1}J_{i,j_0}-\mu,
\end{equation} 
which will occur with probability
\begin{equation}
	p^i_{R,0m}(j_0)=\frac{n_{j_0}}{N}p_{off}(j_0)(P^{(0)}_{mut})^{2},
\end{equation}
where $P^{(0)}_{mut} = (1-p_{mut})^L$ is the probability of no mutations occurring, and its counted twice, once for each offspring. Averaging over all possible  types (of which there are $\Omega = 2^L$) we obtain
\begin{eqnarray}
\bar{\Delta}_{R,0m}	&=& \langle \Delta^i_{R,0m}(j_0)\rangle = \frac{1}{\Omega}\sum_{j_0}\left(\frac{k}{N}J_{ij_0}-\mu\right)\frac{n_{j_0}}{N}p_{off}(j_0)(P^{(0)}_{mut})^{2L}\nonumber\\
	&\mapsto & \left(\frac{k\bar{J}}{N}-\mu\right)\langle p_{off}\rangle_{ext}(1-p_{mut})^{2L},
	\label{no_mut}
\end{eqnarray}
where we have introduced $\bar{J}$, which denotes the strengths $J_{i,j}$ averaged over pairs of interacting extant types and similarly $\langle p_{off}\rangle_{ext}$ denotes the offspring probability average over extant types.\\

{\it Reproduction with 1 mutation} Next we consider the average change to the likelihood function, $H_i$ for type $i$ given that an individual of type $j_0$ reproduces with one copy mutating and ending in $q_o$ and the other not mutating. The change in $H_i$ is given by
\begin{eqnarray}
H_i&\mapsto& 	\frac{k}{N+1}\left[\sum_{j\neq q_0}J_{i,j}n_j + J_{i,q_0}(n_{q_0}+1)\right] - \mu (N+1)\nonumber\\
&=&\frac{k}{N+1}\left(\sum_{j}J_{i,j}n_j-\mu N\right) + \left(\frac{k}{N+1}J_{i,q_0}-\mu\right)\nonumber\\
&=& H_i + \Delta^i_{R,m}(q_0).
\label{1mut} 
\end{eqnarray}
Again we have replaced $N+1$ by $N$ in the first term and introduced the change 
\begin{equation}
	\Delta^i_{R,m}(q_0)= \frac{k}{N+1}J_{i,q_0}-\mu,
\end{equation} 
which will occur with probability  
\begin{equation}
	p^i_{R,m}(j_0)=\frac{n_{j_0}}{N}p_{off}(j_0)p_{j_o \rightarrow q_o},
\end{equation} 
where 
\begin{equation}
p_{j_o \rightarrow q_o} = p_{mut}^{d_{j_{o}q_{o}}} (1-p_{mut})^{L-d_{j_{o}q_{o}}},
\end{equation}
 and $ d_{j_{o}q_{o}}$ is the hamming distance between the sequences $j_o$ and $q_o$,
This means that 
\begin{equation}
\bar{\Delta}^i_{R,1m} = \sum_{j_o q_o} \left(\frac{k}{N+1} J_{i,q_o} -\mu \right)\frac{n_{j_o}}{N}p^{off}_j  p_{mut}^{d_{j_{o}q_{o}}} (1-p_{mut})^{L-d_{j_{o}q_{o}}}.
\end{equation}
By limiting our approximation to the nearest neighbours, and proceeding like in the previous case, we obtain 
\begin{equation}
\bar{\Delta}_{R,1m} = L p^{(o)}_{mut}p^{(1)}_{mut}\left(\frac{k\tilde{J}}{N}-\mu\right)\langle p_{off}\rangle_{ext},
\label{1_mut_ave}
\end{equation}
where $L$ is the number of first neighbours and $P^{(1)}_{mut}=p_{mut}(1-p_{mut})^{(L-1)}$ denotes the probability that exactly one $L$ genes mutate. Notice the difference between $\bar{J}$ introduced in Eq. (\ref{no_mut}) and the averaged quantity $\tilde{J}$ introduced in this equation. The two differs by being averages over different sets of types. Here $\tilde{J}$ is averaged over interaction strengths $J_{ij}$ connecting connecting already occupied type and types hit by a new mutation, \textit{i.e.} types located in the perimeter of the cluster of extant reproducing sites. In contrast $\bar{J}$ is the average of the interaction strength between extant types. We will expect that typically $\tilde{J}<\bar{J}$ because adaptation has favoured mutualistic interactions amongst the extant types. However, an accurate estimate of the two quantities from first principle is of course very difficult.\\

{\it Reproduction with 2 mutations}. Next we consider the average change to the likelihood function, $H_i$ for type $i$ given that an individual of type $j_0$ reproduces with both copies mutating and ending in $q_o$ and $q_1$. The change in $H_i$ is given by
\begin{eqnarray}
H_i&\mapsto & \frac{k}{N+1}\left[\sum_{j\neq q_0,q_1,j_o}J_{i,j}n_j + J_{i,q_0}(n_{q_0}+1)+J_{i,q_1}(n_{q_1}+1)+J_{i,q_1}(n_{q_1}-1)\right]  -\mu (N+1)\nonumber\\
&=&\frac{k}{N+1}\left(\sum_{j}J_{i,j}n_j-\mu N\right) + \left(\frac{k}{N+1}(J_{i,q_0} + J_{i,q_1} - J_{i,j_o})-\mu\right)\nonumber\\
&=& H_i + \Delta^i_{R,2m}(q_0),
\label{2mut}
\end{eqnarray}
where we have consider the fact that the number of individuals of the parent decreases in case of 2 mutations.  Again we have replaced $N+1$ by $N$ in the first term and introduced the change 
\begin{equation}
	\Delta^i_{R,2m}(q_0)= \frac{k}{N+1} ( J_{i,q_0} + J_{i,q_1} - J_{i,j_o})-\mu,
\end{equation} 
which will occur with probability
\begin{equation}
	p^i_{R,m}(j_0)=\frac{n_{j_0}}{N}p_{off}(j_0)p_{j_o \rightarrow q_o}p_{j_o \rightarrow q_1}.
\end{equation} 
And once again limiting our approximation to the nearest neighbour mutations we obtain

\begin{equation}
\bar{\Delta}_{R,2m} = L^2 (P^{(1)}_{mut})^2\left(\frac{k\tilde{J}}{N}-\mu\right)\langle p_{off}\rangle_{ext}.
\label{2_mut_ave}
\end{equation}
\\

Notice the difference between $\bar{J}$ introduced in Eq. (\ref{no_mut}) and the averaged quantity $\tilde{J}$ introduced in this equation. The two differs by being averages over different sets of types. Here $\tilde{J}$ is averaged over interaction strengths $J_{ij}$ connecting already occupied type and types hit by a new mutation, \textit{i.e.} types located in the perimeter of the cluster of extant reproducing sites. In contrast $\bar{J}$ is the average of the interaction strength between extant types. We will expect that typically $\tilde{J}<\bar{J}$ because adaptation has favoured mutualistic interactions amongst the extant types. However, an accurate estimate of the two quantities from first principle is of course very difficult.\\

{\it Killing event} on site $j_0$ leads to 
\begin{eqnarray}
H_i&\mapsto& 	\frac{k}{N-1}\left[\sum_{j\neq j_0}J_{i,j}n_j + J_{i,0_0}(n_{j_0}-1)\right]  -\mu (N-1)\nonumber\\
&=&\frac{k}{N-1}\left(\sum_{j}J_{i,j}n_j-\mu N\right) - \left(\frac{k}{N}J_{i,j_0}-\mu\right)\nonumber\\
&=& H_i - \Delta^i_{R,0m}(j_0). 
\end{eqnarray}
This change occurs with probability $(n_{j_0}/N)p_{kill}$. 

Combining this result with the weighted results in Eqs. (\ref{no_mut}), (\ref{1_mut_ave}) and (\ref{2_mut_ave}) we obtain the following map, which in mean field describes how $\langle H\rangle$ changes as an effect of the microscopic reproduction and killing events
\begin{equation}
	\langle H\rangle \mapsto  \langle H\rangle +A\langle p_{off}\rangle_{ext}	
	-   B p_{kill},
	\label{TaNa_mf_map1}
\end{equation}
	where the coefficients are given by
	
\begin{eqnarray}
	A &=& \left(\frac{k\bar{J}}{N}-\mu\right)(1-p_{mut})^{2L}+\left(\frac{k\tilde{J}}{N}-\mu\right)(P^{(0)}_{mut}+Lp^{(1)}_{mut})Lp^{(1)}_{mut}\label{A_cof}\\
	B &=& \frac{k\bar{J}}{N}-\mu
\end{eqnarray}

We have derived a map for the evolution of $\langle H\rangle$. We now need to close the map, \textit{i.e.} we need a way to express the $H^i$ dependency of $\langle p_{off}\rangle_{ext}$ in terms of $\langle H\rangle$. We could assume
\begin{equation}
	\langle p_{off}(H^i)\rangle_{ext} \mapsto  p_{off}(\langle H\rangle_{ext}). 
	\label{replace1}
\end{equation}    
This procedure gives us the following map for $x_n$ (which we use as shorthand for the iterates of $\langle H\rangle_{ext}$)
\begin{equation}
	x_{n+1}=x_n +A p_{off}(x_n)-B p_{kill}.
	\label{map1}
\end{equation}
The map has a fixed point a $x^*$ given by $p_{off}(x^*)=B p_{kill}/A$. The map is stable if $A<0$ and $B<0$. For $AB<0$ the map is either attractive (repulsive) to the left of $x^*$ and repulsive (attractive) to the right hand  of $x^*$. For $A<0$ and $B<0$ $x^*$ is repulsive in both directions. The conclusion is that the dramatic mean field approximation suggested in Eq. (\ref{replace1}), which corresponds to the replacement $\langle H^n\rangle_{ext} \mapsto \langle H\rangle^n_{ext}$ for all $n\in \mathbb{N}$, wipes out the intermittency. To establish a mean field description of the intermittency we instead expand $p_{off}(H^i)$ in Eq. (\ref{TaNa_mf_map1}) to second order about $x^*$ and replaces only $\langle H^2\rangle_{ext}$ by $\langle H\rangle^2_{ext}$. This leads to a tangent map and we study the intermittency of this map in the next section.
  
\section{Analysis of the map in the neighbourhood of tangency}
We expand $p_{off}(H)$ in Eq. (\ref{TaNa_mf_map1}) to second order about $H^*=\ln [p_{kill}/(1-p_{kill}) ]$,
\begin{equation}
	p_{off}(H) = a_0+a_1(H-H^*)+a_2(H-H^*)^2
	\label{2nd_ord}
\end{equation}
where 
\begin{eqnarray*}
	a_0&=&p_{kill},\\
	a_1&=&p'_{off}(H^*)=p_{kill}(1-p_{kill}),\\
	a_2&=&\frac{1}{2}p''_{off}(H^*)= \frac{1}{2}a_1(1-2p_{kill}).
\end{eqnarray*}
We substitute Eq. (\ref{2nd_ord}) into Eq. (\ref{map1}) and obtain the following map for $\Delta = \langle H\rangle -H^*$
\begin{equation}
	\Delta_{n+1}= b_0 + b_1\Delta_n+b_2\Delta_n^2\equiv f(\Delta_n),
	\label{H_map}
\end{equation} 
where 
\begin{eqnarray*}
	b_0&=&a_0A-Bp_{kill},\\
	b_1&=&1+a_1A,\\
	b_2&=&a_2A.
\end{eqnarray*}

Let $\Delta_c$ be given by  $f'(\Delta_c)=1$ and $\epsilon = f(\Delta_c) - \Delta_c$, \textit{i.e.} at $\Delta_c$ the map has a tangent parallel to the identity and the vertical distance to the identity at this point is $\epsilon$ and is given by
\begin{equation}
	\epsilon = b_0-\frac{(1-b_1)^2}{4b_2}.
\end{equation} 

\begin{figure}%
\includegraphics[width=0.85\columnwidth]{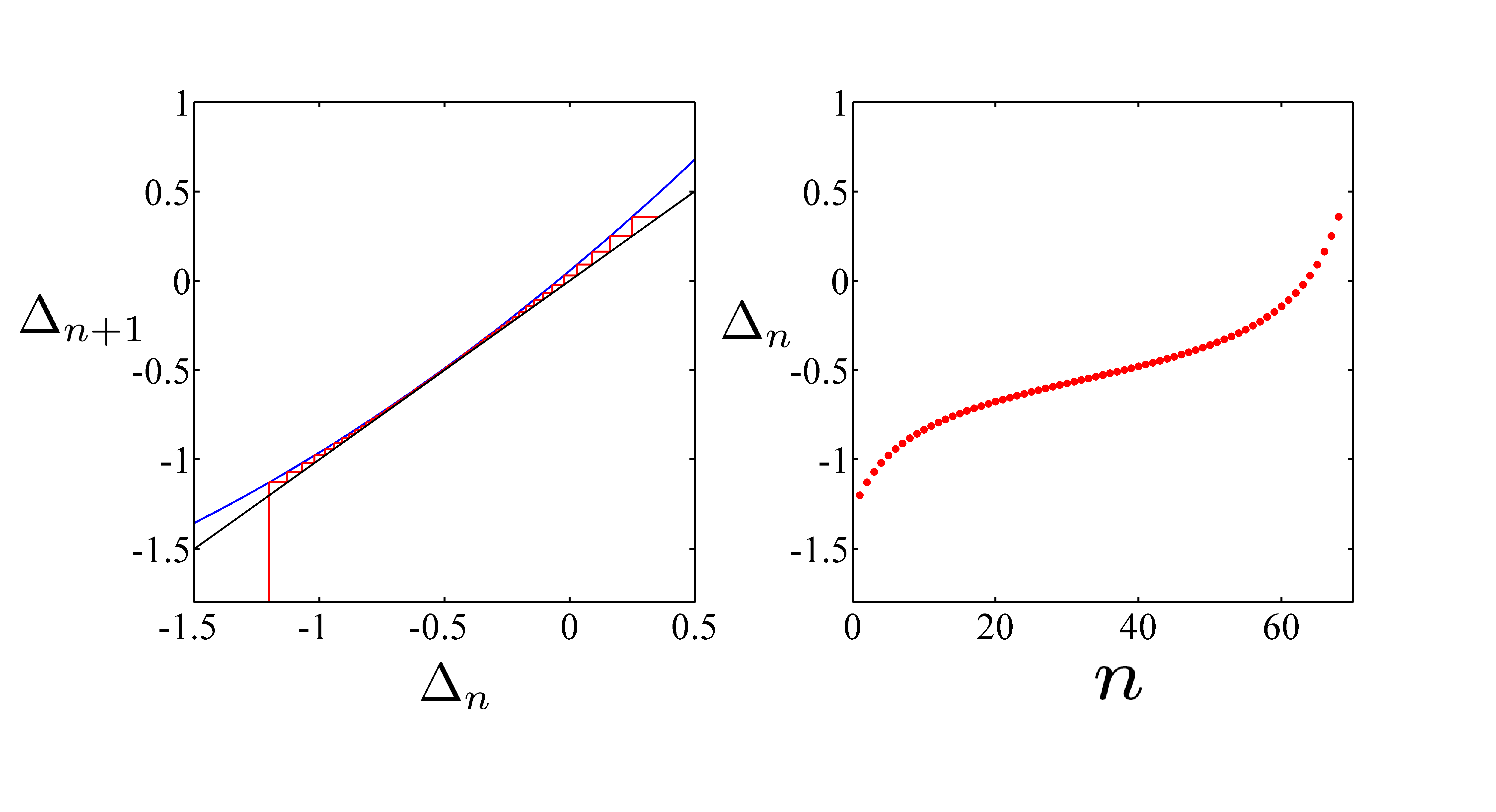}%
\caption{The left panel shows the first $67$ iterations of the map in Eq.(\ref{H_map}), with initial condition $\Delta_0 = -1.2$. The corresponding trajectory is shown in the right panel. The set of parameter values is the same than that as for Fig. \ref{fig:1}, with the corresponding averaged interactions $\bar{J}\approx0.0587$ and $\tilde{J}\approx-0.000001$, thus yielding the coefficients of the map $b_0\approx0.060784$, $b_1\approx 1.167990$ and $b_2\approx 0.151191$.} 
\label{TaNa_H_Map}
\end{figure}

In Figure \ref{TaNa_H_Map} we show an example of an iteration of the map in Eq. (\ref{H_map}) for a set of typical simulation parameters.

The number of iterations $T$ needed to pass through the bottleneck between the map and the identity is of order $T=\pi/\sqrt{\epsilon b_2}$ (see \textit{e.g.} \cite{schuster1} Chap. 4). Hence we have  
\begin{equation}
	\left(\frac{\pi}{T}\right)^2 =b_0b_2-\frac{1}{4}(1-b_1)^2
\end{equation}
We can simplify this expression by by only working to the lowest order in the killing probability and further more we will only include mutation processes considered above, \textit{i.e.} single gene mutations in one or in both offspring. Let us denote by $P_0$ the probability that no mutation occur, \textit{i.e.} $P_0=(1-p_{mut})^{2L}$. Since we neglect all other mutation events than the two kinds just described, we have the approximation 
\begin{equation}
	1- P_0 = (P^{(0)}_{mut}+Lp^{(1)}_{mut})Lp^{(1)}_{mut},
\end{equation}   
in which case Eq. (\ref{A_cof}) becomes
\begin{equation}
	A = \left(\frac{k\bar{J}}{N}-\mu\right)P_0+\left(\frac{k\tilde{J}}{N}-\mu\right)(1-P_0).
\end{equation}
With these approximations we arrive at
\begin{equation}
	\left(\frac{\pi}{T}\right)^2\simeq -\frac{k}{2N}(\bar{J}-\tilde{J})(1-P_0)\left[\frac{k}{N}(\bar{J}-\tilde{J})P_0+\frac{k}{N}\tilde{J}-\mu\right]p_{kill}^2.
\end{equation}
We find that $k(\bar{J}-\tilde{J})N$ is very small and hence that the expression for $(\pi/T)^2$ is well approximated by
\begin{equation}
	\left(\frac{\pi}{T}\right)^2\simeq -\frac{k}{2N}(\bar{J}-\tilde{J})(1-P_0)\left(\frac{k}{N}\tilde{J}-\mu\right)p_{kill}^2.
	\label{T_duration}
\end{equation}

We conclude that our mean field analysis suggests that the length of the qESS, \textit{i.e.} the metastable quiescent epochs, is set by three mechanisms. First the rate of killing. Second the mismatch between the characteristic interaction strength $k\bar{J}$ of the extant types and the carrying capacity as given by the parameter $\mu$ in Eq. (\ref{eq:1}). And thirdly the difference between the typical interaction strength between the extant types, $\bar{J}$ and the typical interaction strength, $\tilde{J}$ across the set of extant and mutant types located in the perimeter of the set of  occupied types. 

It is natural that the duration of the qESS states increases if the rate of killing decreases and it seems also reasonable that the qESS becomes longer if an equilibrium is established between the web of inter-type interactions, as represented by the coupling term in Eq. (\ref{eq:1}), and the demand expressed by the carrying capacity term in the same equation. Finally if the surrounding mutants originating from the extant types experience interactions significantly different from the existing coupling these mutants may very well be able to out compete existing types and thereby destabilise the current qESS. This is what the term $(\bar{J}-\tilde{J})(1-P^{(0)})$ represents.     
 
It is of course interesting to try to relate the prediction for the duration $T$ given by Eq. (\ref{T_duration}) to the actual qESS intermittency observed in simulations of the Tangled Nature model. Unfortunately this turns out not to be straight forward. The problem is that if one simply identify $\bar{J}$ and $\tilde{J}$ by time averages of these quantities during specific qESS periods the right hand side of Eq. (\ref{T_duration})  sometimes ends up being negative. This means that the stochastic dynamics of the Tangled Nature model is not self-averaging. However, if we neglect that the sign of the right hand side can be wrong and simply consider the oder of magnitude of the numerical value predicted for the right hand side the order of magnitude for the number of generations a qESS persists is correct. We consider this as indicating that the average of the couplings restricted to the extant type, $\bar{J}$ and to the couplings averaged over the extant types and those reachable by first generation mutants may very well through Eq. (\ref{T_duration}) yield reasonable results for the durations $T$. However, the mean field theory we have developed does not directly correspond to the time averages of individual qESS.

\section{A consecutive tangent bifurcation model}

Based on the analysis in the previous section we now advance a simple nonlinear dynamical model capable of imitating some features of the macroscopic dynamics that can be typically generated by the
TaNa model. Our model considers families of chaotic attractors in the
vicinity of tangent bifurcations present in low-dimensional iterated maps that 
display intermittency, referred to as intermittency of type I \cite{schuster1}. For convenience these families can be taken from those occurring an infinite number of times in unimodal maps, as
represented by the quadratic logistic map,  $f_{\nu }(x)=1-\nu x^{2}$, $%
-1\leq x\leq 1$, $0\leq \nu \leq 2$.

\begin{figure}[ht!]%
\includegraphics[width=0.85\columnwidth]{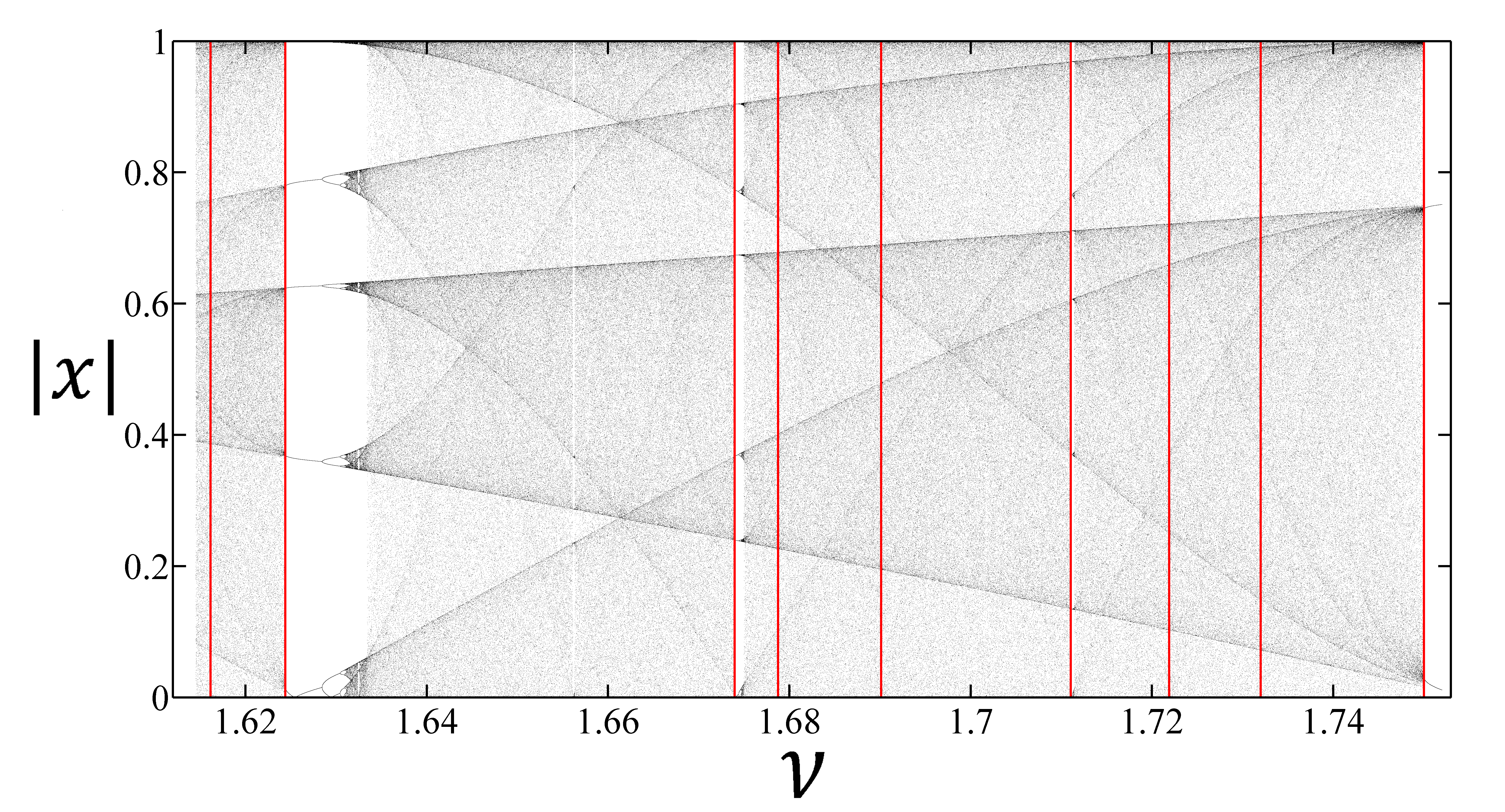}%
\caption{ Families of chaotic attractors with interspersed periodic attractor windows for the logistic map with positions in absolute values, for a range of control parameter values $\nu$. The red lines indicate the control parameter values that correspond to the segments appearing in Fig. \ref{blue_fig}. The periods associated with the vertical red lines are, from left to right, 5, 7, 12, 9, 8, 10, 11 and 3. }
\label{black_fig}%
\end{figure}

Unimodal maps share self-similar families of attractors such as chaotic
attractors that consist of $2^{k}$, $k=0,1,2,\ldots $, bands. For given $k$
the $2^{k}$-band attractors appear for a given range of the control parameter
$\nu$, with the exception of some smaller intervals where periodic attractors
reappear. See Fig. \ref{black_fig}. These intervals are control parameter windows of
regular behaviour that start at a tangent bifurcation at which chaotic
dynamics transforms sharply into periodic motion. The dynamics at the
chaotic attractors in the vicinity of the left edge, $\nu \lesssim \nu _{\tau}$
of the window of periodic attractors, the location $\nu =\nu _{\tau}$ of the
tangent bifurcation, displays intermittency, \textit{i.e.}, at $\nu\lesssim\nu_\tau$ where $\nu_\tau$ is the location of the tangent bifurcation. That is, the map trajectories consist of quasi-periodic
motion interrupted by bursts of irregular behaviour. The iteration time
duration of the quasi-periodic episodes increases as the tangent bifurcation
is approached and the statistical features of these durations have been
shown to display characteristics of various types of noise \cite{schuster2}.
At the tangent bifurcation the duration of the episodes diverges and the
motion becomes periodic. The opening periods $\tau$ of the windows follow the
Sharkovskii ordering \cite{schroeder1}.

The following procedure, which incorporates the criteria identified above for the duration of the qESS, can be used to generate successive quasi-periodic events of different (quasi)-periods mediated by brief erratic bursts, each event associated with a different periodicity $\tau_{n}$ and of different
duration $T_{n}$. First choose a control parameter value $\nu _{0}$ just
left of a window of periodicity $\tau_{0}$ of the logistic map with tangent
bifurcation at $\nu _{\tau_{0}}$, $\delta \nu _{0}\equiv \nu _{0}-\nu
_{\tau_{0}}\lesssim 0$. When the map trajectory with initial condition $x_{0}$
comes out of the bottlenecks formed by $f^{(\tau_{n})}(x)$ and the identity
line (see Fig. \ref{TaNa_H_Map}) to experience a chaotic burst before it is re-injected close to the bottlenecks. The map trajectory evolves in this environment (performing one or more holdup passages and re-injections) until a set of two stochastic conditions is fulfilled, in which case another control parameter value $\nu_1$ is generated just left of a window of periodicity 
$\tau_{1}$ with $\delta \nu _{1}\equiv \nu _{1}-\nu _{\tau_{1}}\lesssim 0$, and so
on for $n=2,3,\ldots $.  These two conditions refer to exceedances associated with two random variables $\delta_1$ and $\delta_2$, distributed by a uniform and a normal distribution, respectively. The conditions are  $\delta_1>\Gamma_1$ and $\delta_2>\Gamma_2$ where  $\Gamma_1$ and $\Gamma_2$ are two prescribed thresholds. Only when the two thresholds are overcomed simultaneously the control parameter value is changed to that of a different window, otherwise the trajectory remains close to the same window. The two implemented thresholds correspond to critical values of the imbalances referred after Eq. (\ref{T_duration}),

\begin{equation}
 \delta_1 = \frac{\frac{k}{N}\tilde{J}-\mu}{p^2_{kill}(1-P_0)} \\ \nonumber
\end{equation}
and
\begin{equation}
 \delta_2 = \frac{\frac{k}{2N}(\bar{J}-\tilde{J})}{p^2_{kill}(1-P_0)}. 
\end{equation}

\begin{figure}[ht!]%
\includegraphics[width=0.85\columnwidth]{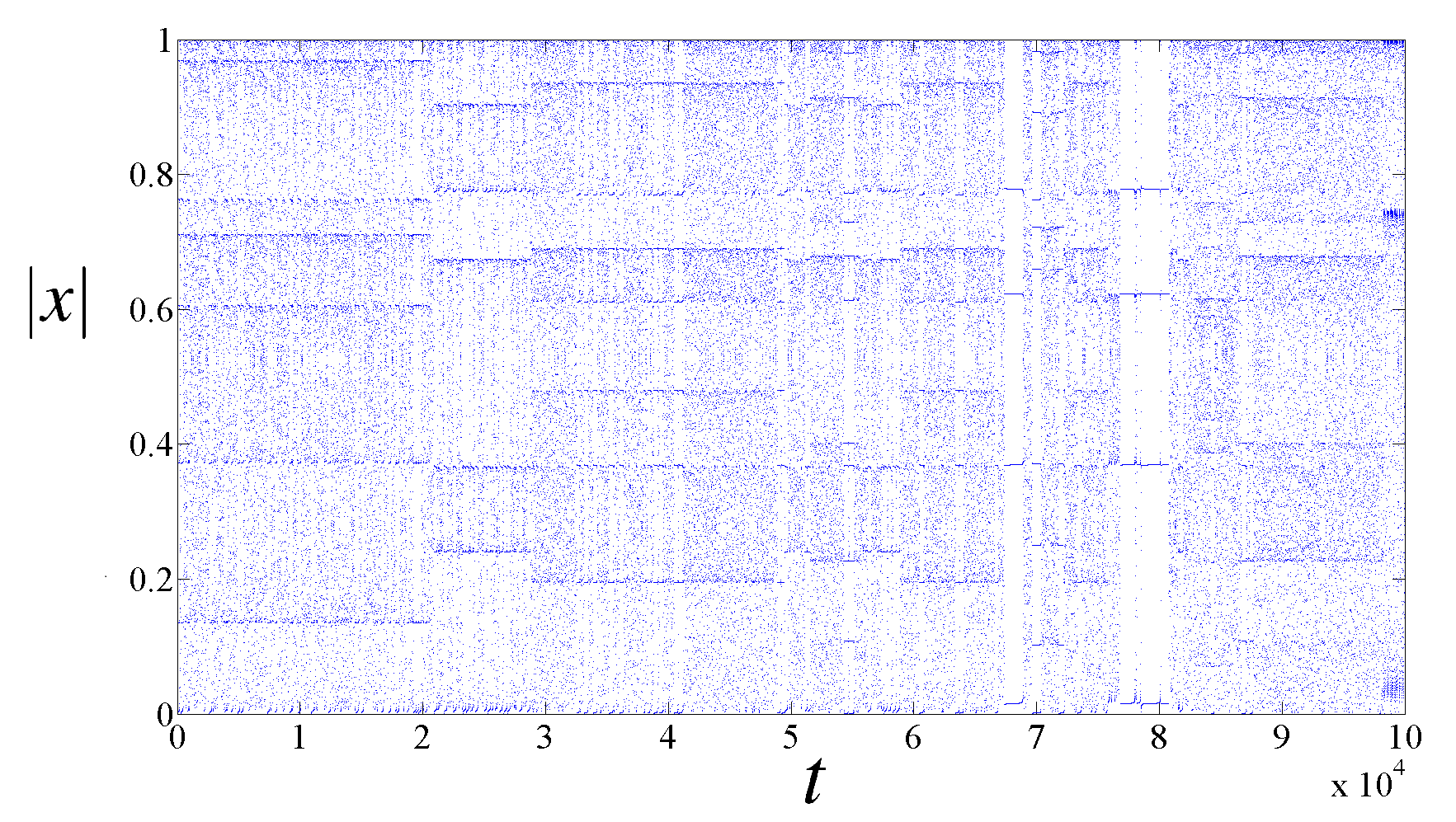}%
\caption{Iterated time evolution of a trajectory generated by the consecutive tangent bifurcation model. The figure is composed of segments, each of which corresponds to a fixed value of the control parameter close to a tangent bifurcation, associated with a given period. Within each segment, many laminar episodes occur separated by chaotic bursts. The periods of the segments are consecutively, from left to right, 8, 7, 9,...  Positions appear in the figure in absolute values.}%
\label{blue_fig}%
\end{figure}

Depending on the threshold values one obtains different dynamical patterns. When of the values of  $\Gamma_1$ and $\Gamma_2$ are small only one or at most a few bottleneck passages take place before there is a change of periodic window. When these values are larger the number of bottleneck passages is large before there is a change in periodic window. This means that the system is sensitive to the imbalances represented by $\delta_1$ and $\delta_2$ and this sensitivity leads to evolutionary changes. When the values of  $\Gamma_1$ and $\Gamma_2$ are large the number of bottleneck passages is large before a change of periodic window takes place. In this case the system is robust to environmental variations. The repetition of this prescription leads to the dynamical behavior shown in Fig. \ref{blue_fig} that can be compared with that obtained
from the TaNa model in Fig. \ref{fig:1}. The quasi-periodic episode of period $\tau_{n}$
is identified with the quasi stable co-existence of $n$ species for a time
period $T_{n}$ in the TaNa model and the chaotic burst at its ending leads
to some extinctions and new mutated species of the following quasi-stable configuration.

This approach can be considered to be a phenomenological modelling of the
original TaNa model. The threshold selections of the periodic windows $\tau_{n}$
at $\nu _{\tau_{n}}$ and of the value of the control parameter distance $\delta
\nu _{n}$ from the corresponding tangent bifurcation can be further
elaborated, \textit{e.g.} by devising specific rules suggested by ecological
principles associated with reproduction, mutation and death, and in this way
obtain a closer reproduction of the dynamics of the TaNa model. Interestingly,
an average decrement of  the variables $\delta \nu_{n}$ with increasing time $t$, that
implies an average increment of the duration of quasi-stable episodes $T_{n}$
with $t$, observed in the TaNa dynamical properties, signals an approach to
the intermittency transition out of chaos. Our modelling by means of the
dynamics associated with families of tangent bifurcations implies (in a
well-defined manner restricted to deterministic nonlinear dynamics) that
the ecological evolution model operates near the onset of chaos, in our case, at
nearly vanishing Lyapunov exponent.

\section{Discussion and Conclusion}   
To explore a possible relationship between the dynamical properties of low-dimensional nonlinear systems and the high-dimensional, and often stochastic, dynamics of relevance to complex systems it is necessary to identify a few robust macroscopic degrees of freedom of the latter that capture the salient features of its dynamical evolution. We have performed such an analysis for a particularly high-dimensional and particularly strongly stochastic model, namely, the Tangled Nature model of evolutionary ecology. We find that despite of the dramatic approximations involved in establishing a local one-dimensional map we nevertheless obtain meaningful and interesting statements concerning the duration of the long quiescent epochs of the TaNa model. 

We then went on to describe how a simple quadratic map is able to reproduce structures that qualitatively exhibit, with a high degree of similarity, the full high-dimensional stochastic model. We advanced a nonlinear dynamical system model consisting of consecutive one-dimensional chaotic attractors near tangent bifurcations that generate time evolution patterns resembling those of the TaNa model in macroscopic scales. These tangent bifurcations are chosen from the infinite families that occur at the onset the periodic windows in the logistic map according to a threshold prescription based on the previously identified mechanisms that control the duration of the basic quasi-stable event generated by the local map derived from the TaNa model. These mechanisms involve imbalances between (average) values of parameters with ecological meaning that define the TaNa model. See Eq. (\ref{T_duration})and text below it.

The generally unanticipated link we established between the macroscopic dynamics of a high–dimensional stochastic model and the intermittent dynamics of low-dimensional systems requires a closer examination. This can be developed, first, by deriving under less sweeping approximations the collapse of degrees of freedom that leads to this correspondence. This would include the derivation from the TaNa model of a more complete, non-local structure, of the map, see \textit{e.g.} Eq. (\ref{H_map}), or coupled maps, that incorporate re-injection mechanisms. And, secondly, by extending the consecutive tangent bifurcation model to fit more closely the mutual interactions that define the TaNa model of evolutionary ecology. The occurrence of the connection between high-dimensional and low-dimensional dynamical model systems offers a new path to the study of complex systems.



 
\bigskip

\bigskip AR and AD-R acknowledges support from DGAPA-UNAM-IN103814 and CONACyT-CB-2011-167978 (Mexican Agencies). HJJ was supported by the European project CONGAS (Grant FP7-ICT-2011-8-317672).
 

\end{document}